\documentclass[final,5p,fleqn]{elsarticle}
\usepackage{amsmath}
\usepackage{amssymb}
\usepackage{graphics}
\usepackage{slashed}
\usepackage{graphicx}
\usepackage[usenames]{color}

\def\simge{\mathrel{%
       \rlap{\raise 0.511ex \hbox{$>$}}{\lower 0.511ex \hbox{$\sim$}}}}
\def\simle{\mathrel{
       \rlap{\raise 0.511ex \hbox{$<$}}{\lower 0.511ex \hbox{$\sim$}}}}

\newcommand \beq{\begin{eqnarray}}
\newcommand \eeq{\end{eqnarray}}

\usepackage[colorlinks=true,linktocpage=true,linkcolor=blue,citecolor=blue]{hyperref}

\usepackage{color}
\usepackage[normalem]{ulem}
\renewcommand\sout{\bgroup \color[rgb]{0.55,0.00,0.99} \ULdepth=-.5ex \ULset}

\begin{document}

\begin{frontmatter}

\title{Constraining kaon PDFs from Drell-Yan and $J/\psi$ production}





\author[a]{Wen-Chen Chang\fnref{email1}}
\fntext[email1]{changwc@phys.sinica.edu.tw}

\author[b,c]{Jen-Chieh Peng\fnref{email2}}
\fntext[email2]{jcpeng@illinois.edu}

\author[d]{Stephane Platchkov\fnref{email3}}
\fntext[email3]{Stephane.Platchkov@cern.ch}

\author[e]{Takahiro Sawada\fnref{email4}}
\fntext[email4]{sawada@icrr.u-tokyo.ac.jp}

\address[a]{Institute of Physics, Academia Sinica, Taipei 11529,
  Taiwan}

\address[b]{Department of Physics, University of Illinois at
  Urbana-Champaign, Urbana, IL 61801, USA}

\address[c]{Department of Physics, National Central University,
  Chung-Li, 32001, Taiwan}

\address[d]{IRFU, CEA, Universit\'{e} Paris-Saclay, 91191
  Gif-sur-Yvette, France}

\address[e]{Institute for Cosmic Ray Research, The University of
  Tokyo, Gifu 506-1205, Japan}

\begin{abstract} 
The kaon parton distribution functions (PDFs) are poorly known due to
paucity of kaon-induced Drell-Yan data. Nevertheless, these Drell-Yan
data suggest a softer valence $u$ quark distribution of the kaon
compared to that of the pion. We discuss the opportunity to constrain
the kaon PDFs utilizing the existing kaon-induced $J/\psi$ production
data. We compare the $K^- / \pi^-$ and $K^+ / \pi^+$ cross-section
ratio data with calculations based on two global-fit parametrizations
and two recent theoretical predictions for the kaon and pion PDFs, and
test the results with two quarkonium production models. The $K^- /
\pi^-$ cross-section ratio for $J/\psi$ production provides independent evidence of different valence quark distributions in pion
and kaon. The $K^+ / \pi^+$ $J/\psi$ data are found to be sensitive to
the gluon distribution in kaon. We show that these $J/\psi$ production
data provide valuable constraints for evaluating the adequacy of
currently available sets of kaon PDFs.
\end{abstract}



\end{frontmatter}


The discovery of the partonic structures of nucleons in deep inelastic
scattering (DIS) has led to extensive theoretical and experimental
advances in our knowledge of the parton distribution functions (PDFs)
in the proton. While the internal structures of the lightest mesons,
the pion and the kaon, are of intense theoretical interest due to
their dual roles as Goldstone bosons and quark-antiquark bound states,
the corresponding experimental information is scarce.  Recently,
significant theoretical efforts have been devoted to the calculations
of the quark and gluon distributions of the lightest mesons based on
Lattice QCD~\cite{Lin:2020ssv, Fan:2021bcr, Salas-Chavira:2021wui} and
various theoretical approaches~\cite{Horn16, Aguilar:2019teb, DSE,
  Watanabe:2017pvl, Lan:2019rba, Han21}. The partonic structures of
mesons are also important for understanding the mass decomposition of
hadrons~\cite{Ji95, Roberts21}.

The early pion-induced Drell-Yan data from CERN and
Fermilab~\cite{NA10DY, E326DY, E615DY} form the basis for extracting
the valence quark distribution of the pion~\cite{Owens, ABFKW, GRV,
  GRS, SMRS}, while the sea-quark and gluon distributions are poorly
determined from these data. Recently, the importance of pion-induced
$J/\psi$ data for constraining the quark and gluon distributions of
pion was suggested~\cite{Peng17, Chang20, Hsieh21, Chang23}, leading
to a new extraction of the pion PDFs from a global fit of pion-induced
Drell-Yan and $J/\psi$ production data in the statistical model
approach~\cite{Bourrely22}.

The kaon PDFs are practically unknown experimentally, since the
$K^-$-induced Drell-Yan data have only been measured by the NA3
collaboration with a limited statistical accuracy~\cite{NA3DY}.
Nevertheless, these data provide the evidence that the valence $\bar
u$ quark distribution of $K^-$ is softer than that of $\pi^-$.  This
difference between the pion and kaon valence quark distributions is
attributed to the breaking of the flavor SU(3) symmetry, resulting in
a larger fraction of kaon's momentum being carried by the $s$ quark
than by the lighter $\bar u$ quark. Further experimental inputs to access
the valence quark as well as the gluon distribution of the kaon, are
of much interest.

In this Letter, we investigate how the existing kaon-induced $J/\psi$
production data can constrain kaon's valence quark and gluon
distributions. In particular, the $K^- / \pi^-$ cross-section ratio
data for $J/\psi$ production provide independent experimental evidence
of a softer $\bar u$ quark distribution of $K^-$ than that of
$\pi^-$. We also show that the $K^+ / \pi^+$ ratio is sensitive to the
gluon distribution of the kaon.

\begin{figure}[tb]
\includegraphics[width=\linewidth]{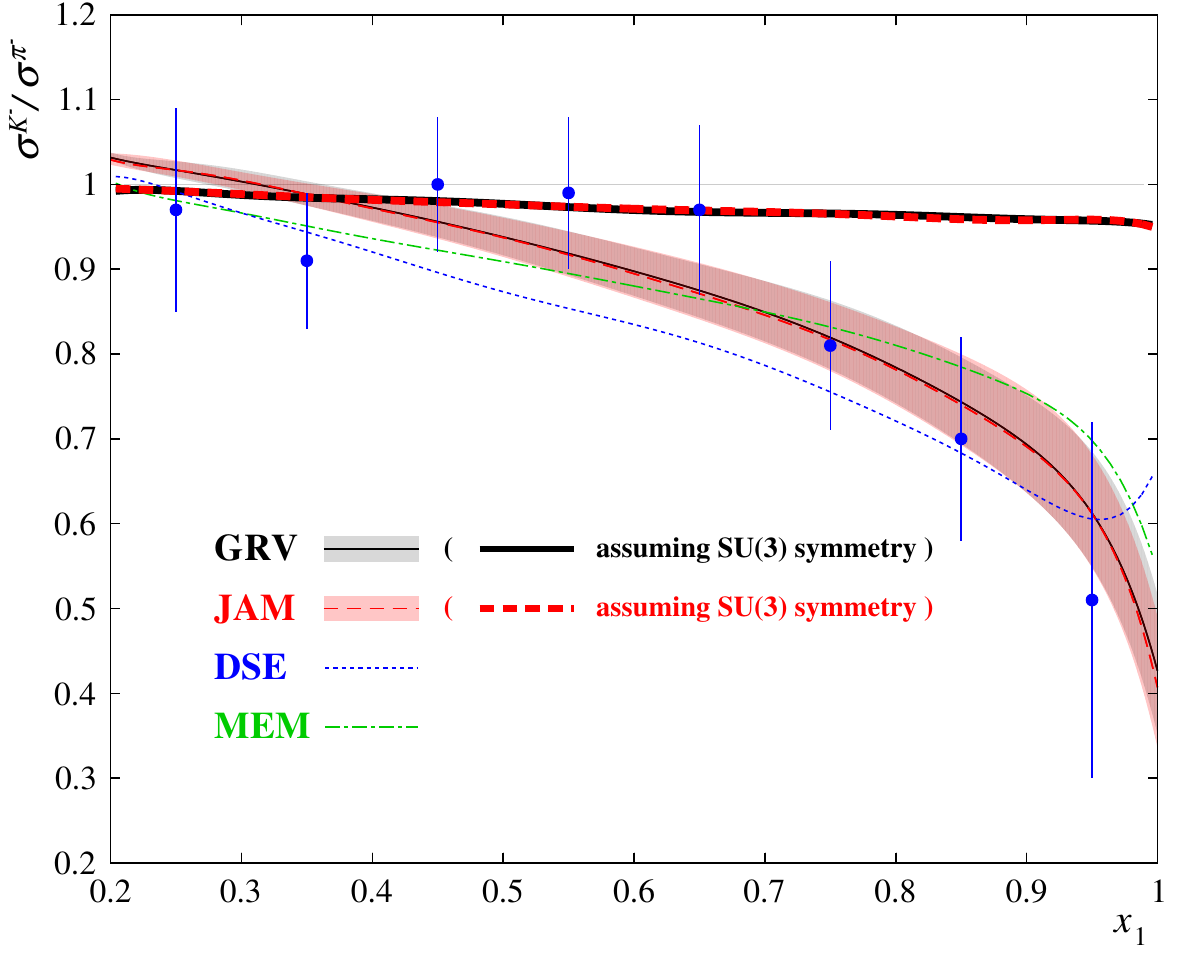}
\caption{Ratios for $K^-$ and $\pi^-$-induced Drell-Yan cross section
  as a function of the momentum fraction $x_1$ on a platinum target
  with 150 GeV beams~\cite{NA3DY}. The solid black, dashed red, dotted
  blue and dot-dashed green curves are NLO Drell-Yan calculations
  using the GRV, JAM, DSE and MEM meson PDFs, respectively. The kaon
  PDFs for GRV and JAM are constructed using the expressions in
  Eqs.(\ref{eq:eq1}) and (\ref{eq:eq2}). The overlapping black and red
  bands denote the uncertainty range of $\kappa$. In addition, two
  thickened curves are the calculations using the GRV (solid black
  curve) and JAM (dashed red curve) meson PDFs assuming
  SU(3)-symmetric distributions for the pion and kaon.}
\label{fig1}
\end{figure}

The only available kaon-induced Drell-Yan data relevant for
constraining the kaon PDFs were collected by the NA3
collaboration~\cite{NA3DY}. The $K^-/\pi^-$ cross-section ratios were
obtained from simultaneous measurements of the $K^- + \rm{Pt} \to
\mu^+ \mu^- + X$ and $\pi^- + \rm{Pt} \to \mu^+ \mu^- + X$ reactions
at 150 GeV. Figure~\ref{fig1} shows the Drell-Yan $K^-/\pi^-$ ratio as
a function of $x_1$, the fraction of the beam momentum carried by the
interacting parton, for the dimuon events with mass $M$ satisfying
$4.1 \le M \le 8.5$ GeV. The fall-off of the ratio at large $x_1$ was
interpreted by NA3 as evidence that the $\bar u$ distribution in
kaon is softer than that in pion~\cite{NA3DY}.


\begin{figure*}[htb]
\centering
\includegraphics[width=0.9\linewidth]{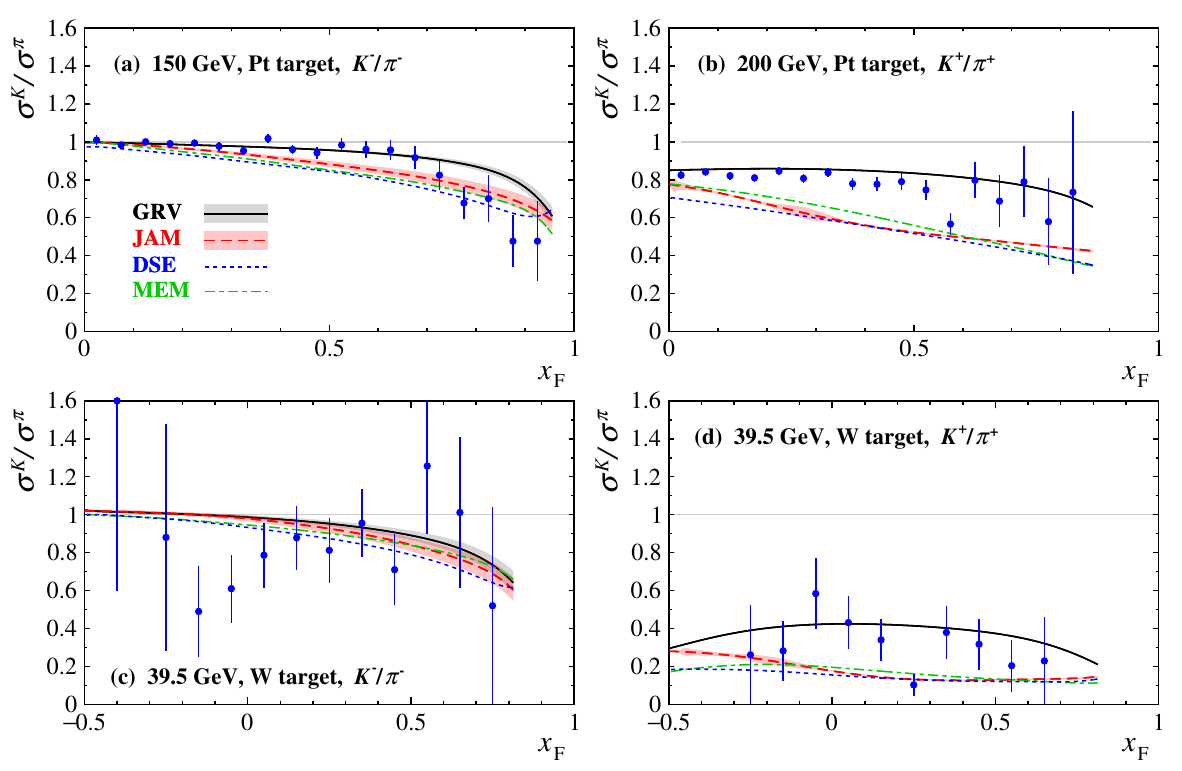}
\caption{Cross-section ratios for $J/\psi$ production versus $x_F$:
  (a) $K^-/\pi^-$ at 150 GeV on a platinum target~\cite{NA3JPSI}, (b)
  $K^+/\pi^+$ at 200 GeV on a platinum target~\cite{NA3JPSI}, (c)
  $K^-/\pi^-$ at 39.5 GeV on a tungsten target~\cite{WA39}, and (d)
  $K^+/\pi^+$ at 39.5 GeV on a tungsten target~\cite{WA39}. The data
  are compared with NLO CEM calculations using various meson PDFs,
  denoted by the solid black, dashed red, dotted blue and dot-dashed
  green curves for the GRV, JAM, DSE and MEM meson PDFs,
  respectively. The black band denotes the uncertainty range of
  $\kappa$ for GRV PDFs while the red one is the combined uncertainty
  of $\kappa$ and the PDF uncertainty for JAM PDFs.}
\label{fig2}
\end{figure*}

In comparison with the Drell-Yan process, the significantly larger
$J/\psi$ production cross sections allow for measurements with much
higher event rates. The NA3 collaboration reported a measurement of
$K^- / \pi^-$ ratios on a platinum target at 150 GeV~\cite{NA3JPSI}
(Fig.~\ref{fig2}(a)). The data covered a broad range in $x_F$
($x$-Feynman) with good statistical accuracy. A comparison between
Fig.~\ref{fig2}(a) and Fig.~\ref{fig1} shows a striking similarity --
while the $K^-/\pi^-$ ratio approaches unity in the region of $x_F <
0.6$, it drops significantly as $x_F$ increases. This similarity
suggests a common origin for the pronounced drop at large $x_1$
($x_F$) for the Drell-Yan ($J/\psi$) $K^-/\pi^-$ cross-section ratios.

The NA3 collaboration also measured the $K^+ / \pi^+$ ratios for
$J/\psi$ production on a platinum target at 200 GeV~\cite{NA3JPSI}
(Fig.~\ref{fig2}(b)). Some differences between the $K^+ / \pi^+$ and
the $K^- / \pi^-$ ratios are noted when comparing Fig.~\ref{fig2}(b)
with Fig.~\ref{fig2}(a) -- while there is a pronounced drop at forward
$x_F$ for the $K^- / \pi^-$ ratio, no such drop is observed for the
$K^+ / \pi^+$ ratio. Moreover, the $K^+/\pi^+$ ratios over the region
$0.0 < x_F < 0.7$ are $\sim$ 20\% lower than the $K^-/\pi^-$
ratios. As discussed below, these differences suggest that the
$K^+/\pi^+$ and $K^-/\pi^-$ ratios are sensitive to different aspects
of the kaon PDFs.

The only other measurement for kaon-induced $J/\psi$ production was
performed by the WA39 collaboration using a 39.5 GeV beam on a tungsten
target~\cite{WA39}.  Both the $K^- / \pi^-$ and the $K^+ / \pi^+$
$J/\psi$ ratios were measured, as shown in Fig.~\ref{fig2}(c) and
Fig.~\ref{fig2}(d), respectively. The $K^+/\pi^+$ ratios at 39.5 GeV
lie significantly lower than those at 200 GeV. As discussed later, the
striking energy dependence of the $K^+/\pi^+$ $J/\psi$ ratios reflects
the difference in the dominant process for $J/\psi$ production at
these two beam energies.

In order to calculate the $K/\pi$ cross-section ratios consistently,
we select theoretical approaches that provide both pion and kaon
PDFs. The earliest attempt was made by Gl\"{u}ck, Reya and Stratmann
(GRS)~\cite{GRS}, who obtained the pion PDFs using the constituent
quark model. To account for the drop of the $K^-/\pi^-$ Drell-Yan
ratios at large $x_1$, GRS~\cite{GRS} proposed the following relations
between the kaon and the pion valence-quark distributions:
\begin{eqnarray}
\bar u^K_v (x) = N_u \bar u^\pi_v (x) (1-x)^\kappa,
\label{eq:eq1}
\end{eqnarray}
where $\bar u^K_v (x)$ and $\bar u^\pi_v (x)$ correspond to the
valence $\bar u$ distributions in $K^-$ and $\pi^-$, respectively.
The value of $\kappa$ was found to be 0.17 at the initial scale of
0.34 GeV$^2$ for NLO calculations. The valence strange-quark
distribution in $K^-$ is assumed to be harder than $\bar u^\pi_v (x)$:
\begin{eqnarray}
s^K_v (x) = 2\bar u^\pi_v (x) - \bar u^K_v (x).
\label{eq:eq2}
\end{eqnarray}
The normalization factor $N_u$ in Eq. (\ref{eq:eq1}), together with
the expression of Eq. (\ref{eq:eq2}), ensures that the following sum
rules for valence-quark distributions in kaon are satisfied:
\begin{eqnarray}
\int_0^1 \bar u^K_v(x) dx =1;~~~~\int_0^1 s^K_v(x) dx =1.
\label{eq:eq3}
\end{eqnarray}
The GRS approach also assumes that the sea-quark and gluon
distributions of the kaon are identical to those of the pion. The GRS
ansatz of deriving the kaon's valence quark distributions is applied
to the GRV~\cite{GRV} and JAM~\cite{JAM} global-fit pion PDFs to
construct the individual corresponding kaon PDFs for the
study. Another approach is the Continuum Schwinger function
Methods~\cite{DSE}, which is a covariant non-perturbative QCD approach
for solving the Dyson-Schwinger Equations (DSE). The final one is the
Maximum Entropy Method (MEM)~\cite{Han21} whose parameters for kaon
PDFs were also obtained from a fit to the NA3 $K^-/\pi^-$ Drell-Yan
data.

To begin, we first compare the NA3 $K^-/\pi^-$ Drell-Yan data with
calculations using four different sets of meson PDFs, namely, GRV,
JAM, DSE, and MEM. The calculations of the next-to-leading-order (NLO)
Drell-Yan cross sections are performed using the DYNNLO
package~\cite{DYNNLO}. The nuclear PDFs, EPPS16~\cite{EPPS16}, were
used for the platinum target, although nuclear effects are expected to
largely cancel in the cross-section ratios.

To illustrate the impact of the NA3 data on the kaon PDFs, we first
present the calculations with GRV and JAM PDFs, by assuming that the
kaon and pion PDFs are related by SU(3) symmetry. As shown by the two
thickened solid black (GRV) and dotted red curves (JAM) in
Fig.~\ref{fig1}, both calculations fail to describe the data in the
region of $x_1 > 0.7$. 

The GRS ansatz is used for constructing the GRV and JAM kaon PDFs. The
best-fit value of $\kappa$ used to modify the valence quark
distribution in Eq.(\ref{eq:eq1}), along with its $1\sigma$
uncertainty, at the scale of Drell-Yan data is determined to be $0.19
\pm 0.04$ by a NLO fit to the NA3 $K^-/\pi^-$ Drell-Yan data. The
uncertainty range of $\kappa$ is denoted by the black and red bands of
modified GRV and JAM PDFs in Fig.~\ref{fig1}. Because an SU(3) flavor
symmetric sea is assumed in these pion PDFs from the global fits, the
sea of the constructed kaon PDFs remains SU(3) flavor symmetric. After
applying the GRS ansatz for the kaon PDFs, the GRV and JAM PDFs could
describe the data nicely, like the DSE and MEM PDFs.

Since the kaon-induced $J/\psi$ data were not included in the
extraction of the above sets of kaon PDFs, it is of great interest to
check how well these various sets of meson PDFs could describe the
$K/\pi$ $J/\psi$ production data.  Such a comparison should provide
additional insight and could help differentiate between these PDF
sets. Unlike the Drell-Yan process, whose production mechanism is well
understood, the precise mechanism for the $J/\psi$ production remains
a topic of active research.  We take two theoretical approaches which
are capable of reproducing many important features of the quarkonium
production in hadron collisions. The first is the color evaporation
model (CEM) at the next-to-leading order~\cite{Gavai,Nelson}, and the
second is the nonrelativistic QCD (NRQCD)
formalism~\cite{Bodwin,Beneke:1996tk}.

Both NLO CEM and NRQCD assume a factorization of the quarkonium
production into hard and soft parts. Perturbative QCD (pQCD) is used
to calculate the short-distance hard part for the production of the $c
\bar c$ pairs in various color and spin states via $GG$, $q \bar q$
and $qG$ subprocesses~\cite{Nason, Mangano}. Motivated by the
quark-hadron duality, the CEM assumes a constant probability, $F$, for
all different $c \bar c$ states with an invariant mass $M_{c \bar c}$
less than the $D \bar D$ threshold, to hadronize into a given
charmonium state. This assumption of a common factor for the
hadronization of different subprocesses greatly reduces the number of
parameters in the CEM. We also assume the same $F$ for pion- and
kaon-induced $J/\psi$ production.  For the production of the
charm-quark pair, we utilize the NLO calculations described in
Refs.~\cite{Nason, Mangano}, widely used in the calculation of
heavy-quark production. The final cross sections are obtained by a
convolution of the hard and soft parts with the parton-parton
luminosity of the associated meson and nucleon PDFs~\cite{Chang20}.

Figure~\ref{fig2} compares the $J/\psi$ ratio data with the
calculations using the four sets of meson PDFs. We find that the data
are in excellent agreement with calculation based on the GRV PDFs,
proposed more than two decades ago. The three more recent meson PDFs
give very similar results, but do not agree with the data well. In
particular, they all predict much smaller values of the $K^+/\pi^+$
ratios at 200 and 39.5 GeV. Moreover, they predict faster fall-off
with $x_F$ than the data for the $K^-/\pi^-$ ratios at 150 GeV and
$K^+/\pi^+$ ratios at 200 GeV. Compared to the Drell-Yan $K^-/\pi^-$
ratios data, the $1\sigma$ uncertainty band in the $K^-/\pi^-$ ratios
is slightly reduced due to the dilution from the $GG$ contribution. In
the $K^+/\pi^+$ ratios, the uncertainty band is significantly reduced
due to the absence of valence-valence $q \bar{q}$ contribution to the
$K^+$-induced production. Although not shown in Fig.~\ref{fig1} and
Fig.~\ref{fig2}, we have also performed calculations using the SMRS
meson PDFs~\cite{SMRS} with results comparable to those obtained with
GRV. Similarly, calculations using the xFitter meson
PDFs~\cite{xFitter} are very close to those of the JAM PDFs.

\begin{figure}[htb]
\includegraphics[width=\linewidth]{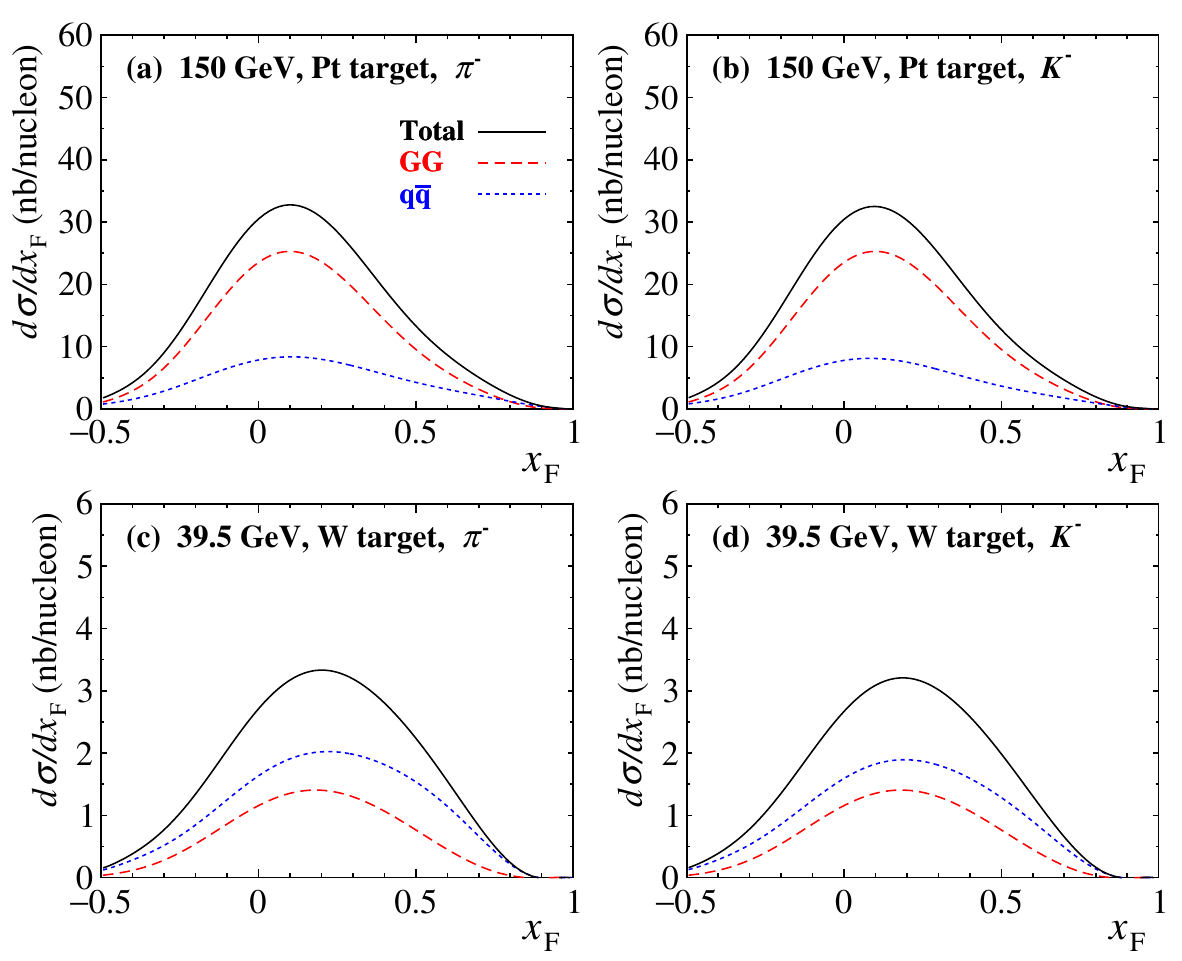}
\caption{Cross sections for $\pi^-$ and $K^-$-induced $J/\psi$
  production calculated using the NLO CEM with the GRV meson PDFs. The
  upper two plots, (a) and (b), are for a beam momentum of 150 GeV on
  a platinum target, and the lower two, (c) and (d), are for a beam
  momentum of 39.5 GeV on a tungsten target. The solid black curves
  are for the total production cross sections, and the dashed red and
  dotted blue curves are the contributions from the $q \bar q$
  annihilation and $GG$ fusion subprocesses, respectively.}
\label{fig3b}
\end{figure}

\begin{figure}[htb]
\includegraphics[width=\linewidth]{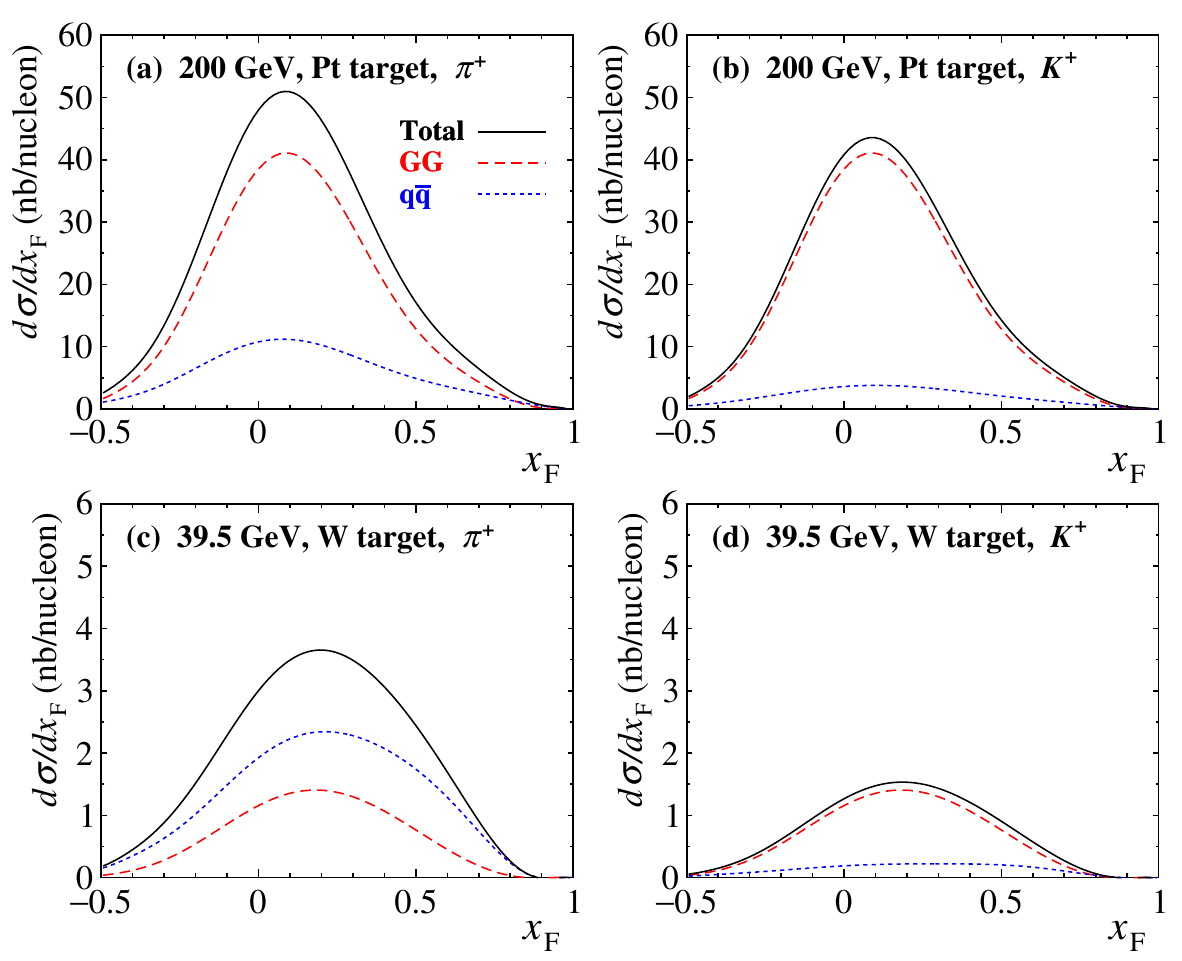}
\caption{Same as Fig.~\ref{fig3b}, but for $\pi^+$ and $K^+$-induced
  $J/\psi$ production.}
\label{fig3a}
\end{figure}

Since the most significant differences between the data and the
calculations in Fig.~\ref{fig2} occur for the $K^+/\pi^+$ ratios of
$J/\psi$ production, it is useful to explore the origin for these
differences. In Figs.~\ref{fig3b} and ~\ref{fig3a}, the calculations
of differential cross sections as a function of $x_F$ for $J/\psi$
production and the individual contributions of the $q \bar q$ and $GG$
channels are shown for the $\pi^-$, $K^-$, $\pi^+$, and $K^+$ beams,
respectively. These results are obtained using the NLO CEM calculation
with the GRV meson PDFs, with the normalization factor $F$ set to
$0.05$~\cite{Chang20}. Both $\pi^-$ and $K^-$ possess $\bar u$ valence
quarks so that the $q \bar q$ contributions to the $J/\psi$ production
cross section are very similar. The $GG$ fusion contributions are the
same in both reactions. Therefore, the $K^-/\pi^-$ ratios are close to
1 except decreasing slightly toward $x_F=1$ due to a softer $\bar
u(x)$ distribution in the kaon. In contrast, the $K^+$ beam contains
$u$ and $\bar s$ valence quarks, which can only annihilate with the
sea quarks in the nucleons, while the $\bar d$ valence quark in
$\pi^+$ can annihilate with the $d$ valence quark in the nucleons
resulting in additional contribution from the $q \bar q$
annihilation. Consequently, the $q \bar q$ contribution is suppressed
compared to the $GG$ fusion for the $K^+$ beam and the $K^+/\pi^+$
ratios turn out to be less than 1 and become very sensitive to the
gluon distribution in the kaon. The prediction of a $K^+/\pi^+$
ratio much lower than the data, as shown in Fig.~\ref{fig2} for the
JAM, DSE, and MEM PDFs, is attributed to their kaon gluon
distributions being smaller than required by the data.

\begin{figure}[tb]
\includegraphics[width=\linewidth]{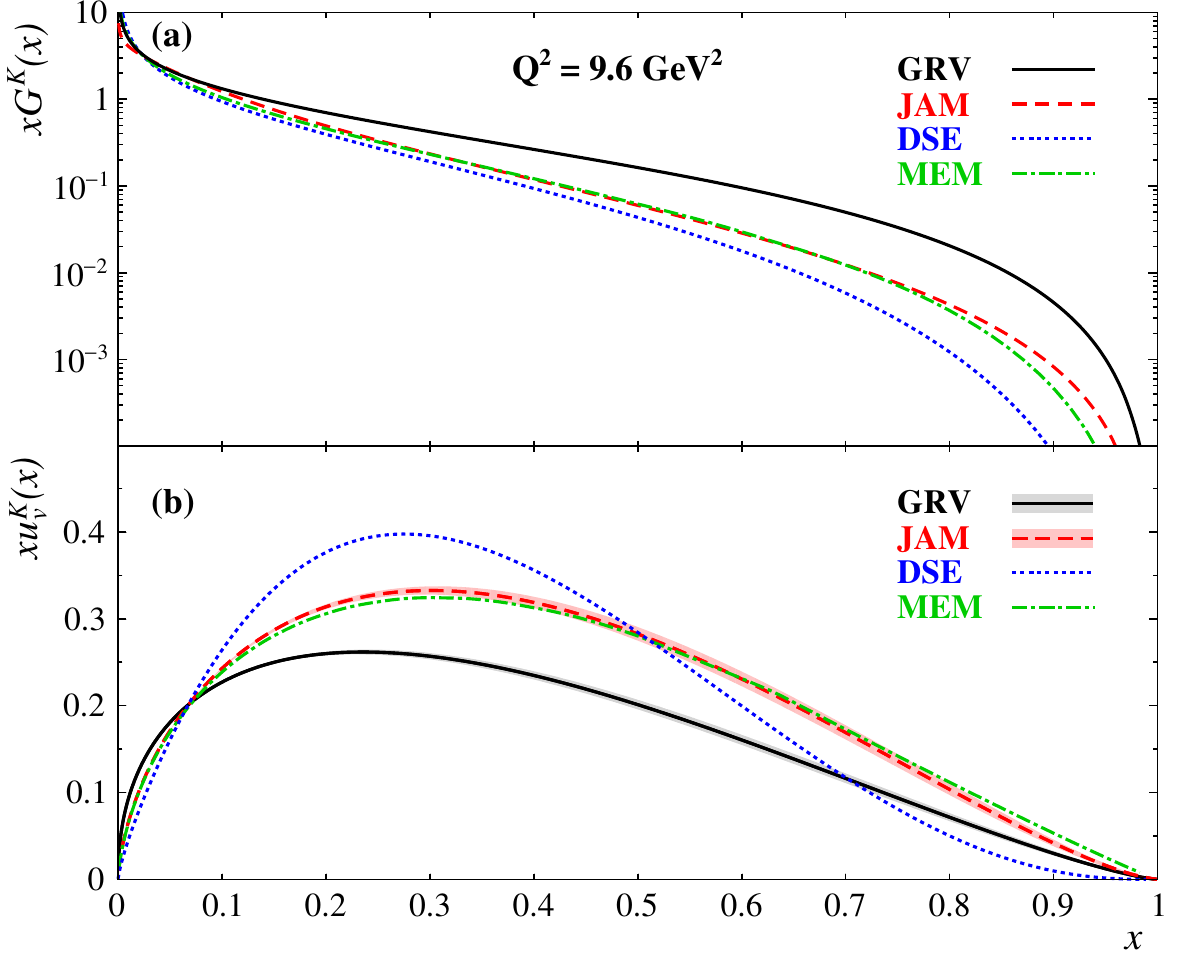}
\caption{Distributions of the gluon ($xG$, (a)) and the valence $u$
  quark ($xu_v$, (b)) for the GRV, JAM, DSE, and MEM kaon PDFs. The
  black and red bands denote the uncertainty range of $\kappa$ for GRV
  and JAM PDFs, respectively.}
\label{fig4}
\end{figure}

Figure~\ref{fig4} shows the gluon ($xG$) and valence $u$ quark
($xu_v$) distributions of the four kaon PDFs at the scale $Q^2=9.6$
GeV$^2$ relevant for $J/\psi$ production. These distributions exhibit
notable differences across the various PDFs. The rapid fall-off at
large $x$ for the JAM, DSE, and MEM gluon distributions in kaon is in
contrast with the much slower drop of the GRV PDF. A behavior similar
to that of the GRV is also observed for the SMRS PDF, not shown in
Fig.~\ref{fig4}.

We note that the pion and kaon gluon PDFs are set to zero at the
initial scale for the DSE~\cite{DSE} and MEM~\cite{Han21} approaches,
i.e., the gluon distribution at large $Q^2$ is solely generated by the
QCD parton radiation process. The gluon radiation from the heavier $s$
quark in kaon is further suppressed with respect to that from the $u$
and $d$ quarks. In contrast, there is already a significant
valence-like gluon distribution at the initial scale for the GRV meson
PDFs~\cite{GRV}.

The $K^+/\pi^+$ ratio data for $J/\psi$ production clearly favor a
harder gluon distribution in pion and kaon than the parametrizations
for the JAM, DSE, and MEM PDFs. This finding is consistent with
observations made in a previous study~\cite{Chang20} that the
pion-induced $J/\psi$ production data favor a gluon distribution in
the pion that is harder than the distributions in JAM and xFitter.

\begin{figure*}[htb]
\centering
\includegraphics[width=0.9\linewidth]{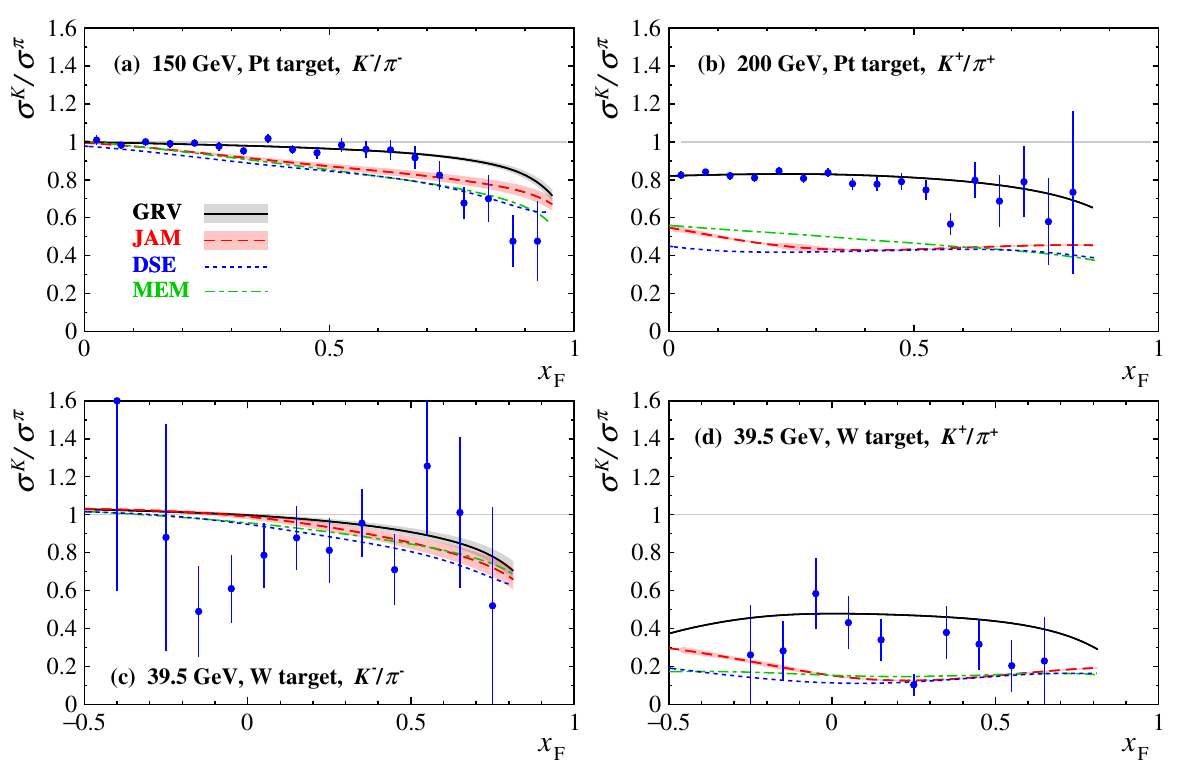}
\caption{Same as Fig.~\ref{fig2}, but for the NRQCD calculations.}
\label{fig5}
\end{figure*}

In order to check whether the evaluation of $K/\pi$ ratios depends on
the model used for the calculation of quarkonium production, we have
also performed calculations using the NRQCD
approach~\cite{Beneke:1996tk}. In NRQCD, the probability of a $c \bar
c$ pair hadronizing into a quarkonium bound state $H$ ($H$ = $J/\psi$,
$\psi(2S)$, or $\chi_{cJ}$) is described by the long-distance matrix
elements (LDMEs), $\langle \mathcal{O}_{n}^{H} [^{2S+1}L_{J}]\rangle$,
depending on the spin, orbital , and total angular momentum quantum
numbers, $S$, $L$ and $J$, respectively, and on the color
configuration ($n$)~\cite{Hsieh21, Chang23, Beneke:1996tk}. These
LDMEs are assumed to be universal and independent of the beam
species. Since the proton PDFs are well determined, the proton-induced
data help in constraining the values of LDMEs common to all charmonium
production data. A satisfactory description of $J/\psi$ and
$\psi^\prime$ production induced by pion and proton beams at
fixed-target energies was recently achieved~\cite{Chang23}. The
extracted LDMEs~\cite{Chang23} are used in the present analysis.

Figure~\ref{fig5} compares the $K/\pi$ cross-section ratio data for
$J/\psi$ production with the calculations performed within the NRQCD
framework using the four meson PDF sets. A comparison of
Fig.~\ref{fig5} and Fig.~\ref{fig2} shows that qualitatively similar
results are obtained for both theoretical approaches. The GRV kaon
PDFs consistently give a better description of the data than the other
three kaon PDFs. The finding that the data favor a harder gluon
distribution in the kaon is also supported by the NRQCD
calculations. The systematic variation of the scale and mass
parameters in our CEM and NRQCD calculations of $J/\psi$ production
cross sections with pions has been extensively studied in
Refs.~\cite{Chang20, Hsieh21, Chang23}.  The preference of the data
for a harder pion gluon distribution remains unchanged for all these
variations. In the current study of $K/\pi$ ratios, systematic
uncertainties from the scale and mass parameters are expected to be
greatly reduced due to cancellation. Other than the uncertainties of
$\kappa$ parameters, we also study the PDF uncertainties of JAM PDFs
since its MC replicas are available. The variations of the $K/\pi$
ratios due to the PDF uncertainties are found to be negligibly small,
compared to those resulting from the uncertainties of the $\kappa$
parameter. A similar study for the other PDFs like GRV, DSE, and MEM
is not possible since the corresponding information on the PDF
replicas is not available.

Although our analysis has focused on the $K/\pi$ ratios for $J/\psi$
production, we note that other experimental observables in
kaon-induced $J/\psi$ production are also of great interest. In
particular, the difference between the $K^-$ and $K^+$-induced
$J/\psi$ production cross sections on an isoscalar target, e.g.,
$\sigma_{J/\psi} (K^- + D) - \sigma_{J/\psi} (K^+ + D)$, can provide a
precise determination of the valence $\bar u$-quark distribution of
the kaon. It can be readily shown that the above cross-section
difference is proportional to the product of the valence $\bar
u$-quark distribution of $K^-$ and the valence quark distribution in
the nucleon. A similar suggestion was considered
earlier~\cite{Londergan:1996vh}, but for the difference of Drell-Yan
cross sections, $\sigma_{DY} (K^- + D) - \sigma_{DY} (K^+ + D)$. The
much larger production cross sections for the $J/\psi$ production than
for the Drell-Yan process could provide an independent,
high-statistics measurement of the valence $\bar u$-quark distribution
of $K^-$ in future kaon-induced $J/\psi$ production
experiments~\cite{AMBER}.

We summarize the main findings of this paper. First, we confirm that
the ansatz proposed by the GRS~\cite{GRS}, namely that the valence
quark distributions for the kaon are related to that of the pion by
Eqs. (\ref{eq:eq1}) and (\ref{eq:eq2}), and that the sea-quark and
gluon distributions of the pion and kaon are identical, can
satisfactorily describe the only existing $K^-/\pi^-$ Drell-Yan ratio
data from NA3. We then note that the $K^-/\pi^-$ $J/\psi$ ratio data
from NA3 provide independent evidence that the $\bar u$ valence quark
distribution of the kaon has a softer $x$ distribution than that of
the pion. The $K^+/\pi^+$ $J/\psi$ ratio data are shown to be very
sensitive to the gluon distribution of the kaon, and can be used to
discriminate the various sets of existing kaon PDFs. In particular,
the $K^+/\pi^+$ ratio data for $J/\psi$ production favor the GRV kaon
PDFs, which have a gluon distribution larger than those obtained by
JAM, DSE, and MEM. The good agreement obtained using the GRV PDFs
would therefore indicate that the $J/\psi$ ratio data are consistent
with the scenario of nearly equal pion and kaon gluon
distributions. These findings illustrate the usefulness of the
$J/\psi$ $K/\pi$ ratio data for constraining the poorly known kaon
PDFs. A first attempt to extract the kaon PDF from these data using
the statistical model was recently reported~\cite{Claude24}.  When new
data on the kaon-induced Drell-Yan and $J/\psi$ production anticipated
at the AMBER experiment~\cite{AMBER}, together with the DIS data based
on the Sullivan process proposed for the China
EIC~\cite{Anderle:2021wcy, Xie:2021ypc} and
U.S. EIC~\cite{Accardi:2012qut, Aguilar:2019teb}, become available,
further refinement in the parameterization of kaon PDFs could be
considered. For example, the requirements of the same valence quark
momentum sum for kaon and pion, as well as the SU(3) flavor symmetry
in the meson seas, could be relaxed in future global fits. Finally,
persistent theoretical efforts to improve our understanding of the
reaction mechanism involved in quarkonium production are of utmost
importance for reducing the uncertainties in extracting the meson
PDFs.

We acknowledge helpful discussions with Craig Roberts, Chengdong Han,
Rong Wang, and Xurong Chen and information they provided on the DSE
and MEM meson PDFs. This work was supported in part by the
U.S. National Science Foundation Grant No. PHY-2210452 and the
National Science and Technology Council of Taiwan (R.O.C.).


\bibliographystyle{elsarticle-num} 

\bibliography{ref}


\end{document}